\documentclass[12pt]{article}
\usepackage{amsmath}
\usepackage{graphicx}
\usepackage{subfigure}
\usepackage{hyperref}

\begin{document}

\author{Ernst Trojan and George V. Vlasov \and \textit{Moscow Institute of Physics
and Technology} \and \textit{PO Box 3, Moscow, 125080, Russia}}
\title{Shock waves in cosmic strings: growth of current}
\maketitle

\begin{abstract}
Intrinsic equations of motion of superconducting cosmic string may admit
solutions in the shock-wave form that implies discontinuity of the current
term $\chi $. The hypersurface of discontinuity propagates at finite
velocity determined by finite increment $\Delta \chi =\chi _{+}-\chi _{-}$.
The current increases $\chi _{+}>\chi _{-}$ in stable shocks but transition
between spacelike ($\chi >0$) and timelike ($\chi <0$) currents is
impossible.
\end{abstract}

\sloppy

\section{Introduction}

Cosmic strings are 2-dimensional topological defects that may be formed at a
phase transition in the early universe \cite{Kibble80,Vilenkin85}. Although
their direct detection is still impossible with modern technical equipment,
they are believed to be responsible for several astrophysical phenomena
associated with gravitational lensing, gravitational waves, particle
acceleration and gamma-ray bursts \cite{BHV2001,FV2006}, and their
theoretical study can give better understanding of the global structure of
the universe.

The dynamics cosmic strings is determined by the Lagrangian $\Lambda $, also
called as equation of state (EOS). The Lagrangian of the Goto-Nambu model $%
\Lambda =-m^2$ depends on the only mass parameter $m$. Cosmic strings may
have non-trivial internal structure, giving rise to additional degrees of
freedom associated, for example, with transversal oscillations of the
worldsheet, or actual particles (bosons or fermions) trapped in the defect
core \cite{W85}. Such current-carrying or ''superconducting'' strings are
described by the Lagrangian $\Lambda =\Lambda (m,\chi )$ which depends on
the magnitude of the current $\chi =-h^{ab}\partial \phi /\partial \sigma
^a\partial \phi /\partial \sigma ^b$ that includes the surface metric $h^{ab}
$ with respect to local coordinates $\sigma ^a$ within the string
worldsheet, and the scalar field $\phi $ coupled to the string.

There are only few explicit analytical models of superconducting strings.
The linear EOS \cite{R2000} $\Lambda =-m^2-\chi /2$ is applied to the
strings that carry fermionic currents. Bosonic current carriers are
described by much more complicated EOS, namely, the ''transonic'' model\ 
\cite{NO87,SPG87,CHT87} 
\begin{equation}
\Lambda =-m\sqrt{m^2+\chi }  \label{lagr2}
\end{equation}
the ''polynomial'' model \cite{Carter2000}: 
\begin{equation}
\Lambda =-m^2-\frac \chi 2\left( 1+\frac \chi {m_{*}^2}\right)  \label{lagr3}
\end{equation}
the ''rational'' model \cite{CP95,LP2009}: 
\begin{equation}
\Lambda =-m^2-\frac \chi 2\left( 1+\frac \chi {m_{*}^2}\right) ^{-1}
\label{lagr4}
\end{equation}
and the ''logarithmic'' model \cite{Carter2000,HC08}: 
\begin{equation}
\Lambda =-m^2-\frac{m_{*}^2}2\ln \left( 1+\frac \chi {m_{*}^2}\right)
\label{lagr5}
\end{equation}
where additional mass parameter $m_{*}$ is introduced.

In contrast to the Goto-Nambu model, the superconducting string models admit
a large variety of stable loop solutions (or, vortons) and their
intercommutation may frequently take place \cite{DS89,LS2003}. The problem
of vorton stability and evolution of superconducting string networks attract
the interest of researchers because such string loops can accumulate
significant part of the universe mass and the expected dominant effect of
their reconnection is particle radiation (when some particles are expelled
away).

The equations of motion of superconducting strings admit solutions in the
form of infinitesimal perturbations of two types \cite{Carter89a}: extrinsic
(transverse) perturbations of the worldsheet that are oscillations of the
string geometry,\ and intrinsic (longitudinal) perturbations within the
worldsheet that are oscillations of the current $\chi $. However, this
equilibrium, or so-called ''elastic'' description is admitted until small
transverse oscillations grow up and form abrupt changes in the geometry
(kinks), or when small longitudinal oscillations are accumulated into
finite-amplitude discontinuities of the current (shock waves) \cite
{MP00,CCMP02}. Such strings loops may tend to fold on themselves and make
contact points of self-intersection so that it will be energetically favored
to some of the trapped particles to move out of the string, thus, generating
emission, associated with possible visible consequences \cite
{Kibble80,Vilenkin85,BHV2001,FV2006}.

The dynamical evolution of a fermionic current carrier can develop only
kinks, while a bosonic current carrier can develop kinks at time-like (or
''electric'') currents $\chi <0$ and shocks at space-like (or ''magnetic'')
currents $\chi >0$ \cite{MP00,CCMP02}\textsc{. }In the latter case nonlinear
effects become dominant rapidly in longitudinal perturbations and a stable
discontinuity appears: the magnitude $\chi $ becomes discontinuous and the
total energy is no longer conserved. This singular behavior of
superconducting strings cannot be assigned to some projection effect because
it is seen in simple analytical analysis \cite{V99} as well as it is
observed in numerical simulations performed both in two dimensions \cite
{MP00} and in three dimensions \cite{CCMP02}. An evident hint to a shock
wave can be also recognized in numerical simulation of a 2-dimensional
vorton in the magnetic regime (Figure 7 in Ref. \cite{BS2008}).

In the present paper we complete the analytical analysis of shocks in
superconducting cosmic strings.\ Particular interest is focused on the
behavior of the bosonic current, the shock velocity, the transition between
electric and magnetic regimes.

\section{Small perturbations and shocks}

The physical behavior of superconducting strings is described by the
equations of motion. Their projection along the string worldsheet yields the
'intrinsic' equations or the conservation laws \cite{Carter89a}\textrm{\ } 
\begin{equation}
\eta _\mu ^\nu \nabla _\nu \,j^\mu =0  \label{cur1}
\end{equation}
for $j^\mu =\mu v^\mu $ and $j^\mu =nv^\mu $ where tensor $\eta _\mu ^\nu
=v^\nu v_\mu -u^\nu u_\mu $ is composed of mutually orthogonal unit vectors (%
$u^\mu u_\mu =-1=-v^\mu v_\mu $ and $u^\mu v_\mu =0$), and dynamical
parameters \cite{CP95} 
\begin{equation}
\begin{array}{c}
\mu ^2=\chi \qquad n^2=K^2\chi \qquad \chi >0 \\ 
\mu ^2=-K^2\chi \qquad n^2=-\chi \qquad \chi <0
\end{array}
\label{x1}
\end{equation}
are determined by function \cite{Carter2000}\ 
\begin{equation}
K=-\frac 12\left( \frac{d\Lambda }{d\chi }\right) ^{-1}  \label{k}
\end{equation}

The longitudinal or 'sound' perturbations run within the worldsheet at
velocity \cite{Carter89a} $c^2=c_L^2=\left( n/\mu \right) d\mu /dn$ that in
the light of (\ref{x1})-(\ref{k}) is written in the form\ 
\begin{equation}
c^2=\left( 1+2\frac \chi K\frac{dK}{d\chi }\right) ^{\mp 1}  \label{soundl}
\end{equation}
where the upper (lower) signs correspond to $\chi >0$ ($\chi <0$). It is
always $K<1$ at $\chi <0$, $K\left( 0\right) =1$ and $K>1$ at $\chi >0$
because $dK/d\chi >0$ for all known models (\ref{lagr2})-(\ref{lagr4}) .

According to the linear EOS of fermionic cosmic strings $\Lambda =-m^2-\chi
/2$, the sound speed (\ref{soundl}) coincides with the speed of light $%
c=c_{light}=1$. The bosonic string models (\ref{lagr2})-(\ref{lagr4}) are
endowed with finite sound speed dispersion $dc/d\chi \neq 0$ that provides a
possibility for development of shock waves \cite{V99}. As a matter of fact,
the fermionic current-carrying strings can develop only kinks, while the
bosonic strings admit both kinks and shocks \cite{MP00,CCMP02}.

The front of a shock wave is a hypersurface in 4-dimensional space whose
equation is given in the form $\phi \left( x\right) =0$. It implies that $%
\phi _{+}\equiv \phi _{-}$ for any two points in the hypersurface but, in
general, $\phi _{+}-\phi _{-}\neq 0$ for arbitrary two points beyond the
hypersureface, particularly, when ''$-$'' and ''$+$'' label the states
before and behind the front. A stable shock front is associated with
characteristic space-like vector 
\begin{equation}
\lambda _\nu \equiv \frac{d\phi }{dx_\nu }  \label{la0}
\end{equation}
which is the same before and behind the front $\lambda _{+\nu }=\lambda
_{-\nu }=\lambda _\nu $. It can be presented in the basis $\left( u_\nu
,v_\nu \right) $ in the following form

\begin{equation}
\lambda _{-\nu }=\frac{w_{-}u_{-\nu }+v_{-\nu }}{\sqrt{1-w_{-}^2}}\qquad
\lambda _{+\nu }=\frac{w_{+}u_{+\nu }+v_{+\nu }}{\sqrt{1-w_{+}^2}}
\label{la1}
\end{equation}
where parameters $w_{-}$ and $w_{+}$ play the role of velocities whose
essence is clarified below.

All dynamical variables in the intrinsic equations of motion (\ref{cur1})
are dependent on function $\phi \left( x\right) $, and, crossing the
discontinuity, we have \cite{A1989} 
\begin{equation}
\eta _{+\mu }^\nu \nabla _\nu j_{+}^\mu =\eta _{-\mu }^\nu \nabla _\nu
j_{-}^\mu +\left[ \eta _\mu ^\nu \nabla _\nu j^\mu \right]  \label{fi2}
\end{equation}
where square brackets imply a jump of parameters across the discontinuity.
In the light of (\ref{cur1}), equation (\ref{fi2}) implies 
\begin{equation}
\left[ \eta _\mu ^\nu \frac{Dj^\mu }{d\phi }\frac{d\phi }{dx^\nu }\right]
=\left[ \lambda _\mu \frac{Dj^\mu }{d\phi }\right] =\lambda _\mu \frac{%
\,j_{+}^\mu -\,j_{-}^\mu }{\phi _{+}-\phi _{-}}=\frac{\lambda _{+\mu
}\,j_{+}^\mu -\lambda _{-\mu }\,j_{-}^\mu }{\phi _{+}-\phi _{-}}=0
\label{f3}
\end{equation}
and at finite $\phi _{+}-\phi _{-}\neq 0$ it yields relation at the front $%
\lambda _{+\mu }\,j_{+}^\mu =\lambda _{-\mu }\,j_{-}^\mu $ , that implies $%
\lambda _{+\mu }\,\mu _{+}v_{+}^\mu =\lambda _{-\mu }\,\mu _{-}v_{-}^\mu $
and $\lambda _{+\mu }\,n_{+}u_{+}^\mu =\lambda _{-\mu }\,n_{-}u_{-}^\mu $.
Taking (\ref{la1}), we find the shock parameters 
\begin{equation}
w_{-}^2=\frac{n_{+}^2}{\mu _{+}^2}\frac{\mu _{+}^2-\mu _{-}^2}{%
n_{+}^2-n_{-}^2}\qquad w_{+}^2=\frac{n_{-}^2}{\mu _{-}^2}\frac{\mu
_{+}^2-\mu _{-}^2}{n_{+}^2-n_{-}^2}  \label{q1}
\end{equation}

\begin{figure}[tbp]
\caption{(a): The shock wave [shaded] propagates at velocity $D_-$, the flow
behind it acquires velocity $D_+$. (b): The shock velocities $w_+$ and $w_-$
are determined in the co-moving reference frame. (c): The joint shock
transition is linking currents $\chi_-<0$ and $\chi_+>0$.}
\label{flo1}{\includegraphics[scale=0.4]{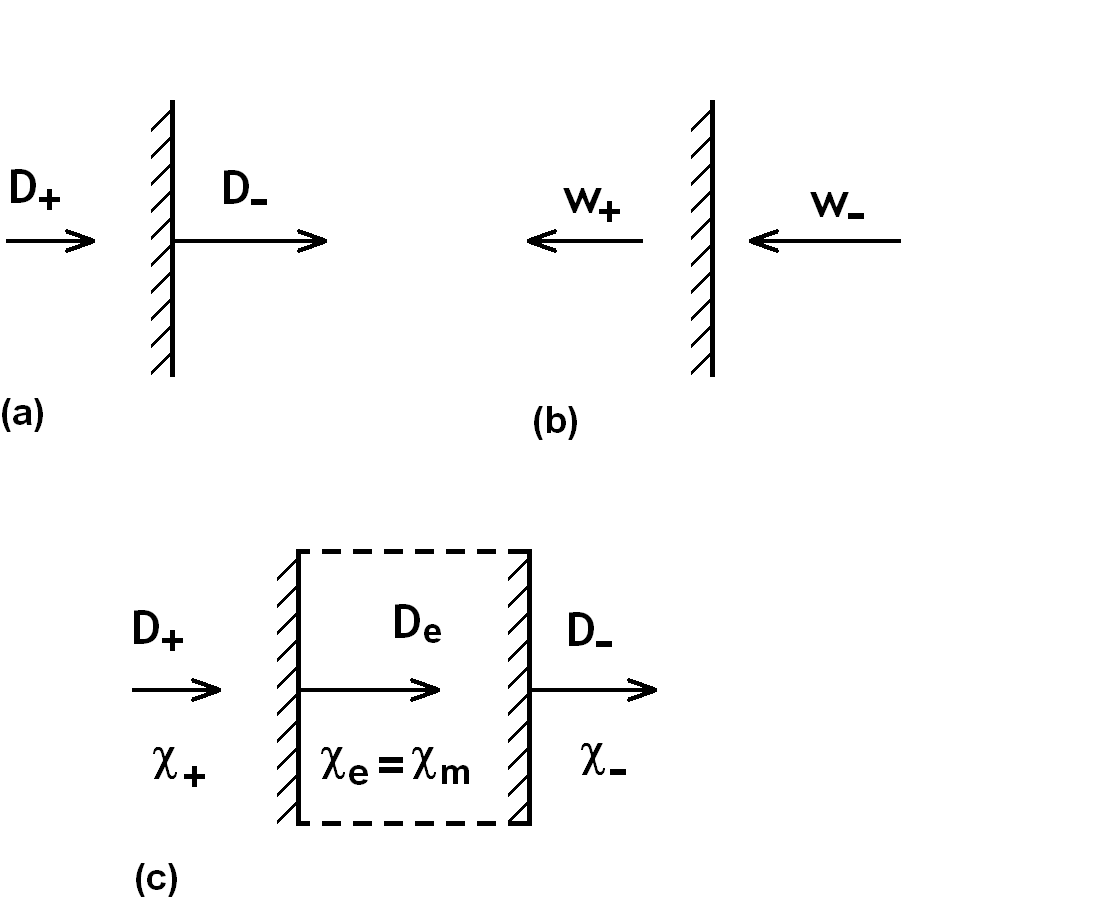}}
\end{figure}

Consider a discontinuity in the the laboratory reference frame (see scheme
in Fig.~\ref{flo1}a) and, altogether, in the reference frame, co-moving its
front (see Fig.~\ref{flo1}b). The front is at rest and the flow before it
has finite velocity, say, $w_{-}$, while the flow behind the front has
velocity $w_{+}$. When we turn to the laboratory reference frame (Fig.~\ref
{flo1}a), the shock-wave propagates along the string at velocity $%
D_{-}=-w_{-}$, while the velocity behind the front $%
D_{+}=(w_{+}-w_{-})/(1-w_{-}\,w_{+})$ is finite if $w_{+}\neq w_{-}$.
Formula (\ref{q1}) yields $w_{-}\rightarrow $ $w_{+}$ $\rightarrow $ $c$ in
the sound limit $n_{+}\rightarrow n_{-}$ and $\mu _{+}\rightarrow \mu _{-}$.
Hence, $D_{-}=-c$\ and $D_{+}=0$ corresponds to a sound wave (\ref{soundl})
because no material flow is generated behind the shock.

Taking $\mu $ and $n$ from equation\ (\ref{x1}), we determine the shock wave
velocity in the magnetic regime ($\chi >0$): 
\begin{equation}
w_{-}^2=w_{+}^2\frac{K_{+}^2}{K_{-}^2}=K_{+}^2\frac{\chi _{+}-\chi _{-}}{%
K_{+}^2\chi _{+}-K_{-}^2\chi _{-}}  \label{w1}
\end{equation}
and in the electric regime ($\chi <0$): 
\begin{equation}
w_{-}^2=w_{+}^2\frac{K_{-}^2}{K_{+}^2}=\frac 1{K_{+}^2}\frac{K_{+}^2\chi
_{+}-K_{-}^2\chi _{-}}{\chi _{+}-\chi _{-}}  \label{w2}
\end{equation}

\section{Growth of current}

Existence of a stable shock is possible when the\textrm{\ }evolutionary
condition is satisfied \cite{Taub78,CF59}. For the shock waves in strings it
implies \cite{V99}: 
\begin{equation}
w_{-}>c_{-}\qquad w_{+}<c_{+}  \label{evol}
\end{equation}
that coincides with the same criterion for the ordinary shock waves.
However, this fact can be predicted immediately by means of very easy
analysis. The number of intrinsic equations of motion (\ref{cur1}) for $%
j^\mu =\mu v^\mu $ and $j^\mu =nv^\mu $ is \textit{two}, while there is only 
\textit{one} material parameter $\chi $: hence, the shock in a string has
exactly \textit{one} degree of freedom. The shock waves in continuous medium
are described by \textit{three} independent equations of motion \cite
{Taub78,CF59}, while there are \textit{two} material parameters (the
pressure and entropy), and, again, there is only \textit{one} degree of
freedom. It implies that the shocks in fluids and cosmic strings obey the
same constraint (\ref{evol}), imposed on their velocities.

Now consider small-amplitude shock waves. If increment $\Delta \chi =\chi
_{+}-\chi _{-}$ is small with respect to $\left| \chi _{\pm }\right| $,
formulas (\ref{w1})- (\ref{w2}) can be expanded in a series of $\Delta \chi $%
. According to (\ref{w1})-(\ref{w2}), we have 
\begin{equation}
\frac{w_{-}^2}{c_{-}^2}-1=1-\frac{w_{+}^2}{c_{+}^2}=c_{-}^2\frac
Q{K_{-}^2}\chi _{-}\Delta \chi \qquad \chi _{-}>0  \label{w11}
\end{equation}
\begin{equation}
\frac{w_{-}^2}{c_{-}^2}-1=1-\frac{w_{+}^2}{c_{+}^2}=-\frac 1{c_{-}^2}\frac
Q{K_{-}^2}\chi _{-}\Delta \chi \qquad \chi _{-}<0  \label{w22}
\end{equation}
where 
\begin{equation}
Q=3\left( \frac{dK}{d\chi }\right) ^2-K\frac{d^2K}{d\chi ^2}  \label{kkkk}
\end{equation}
and $K$ is determined by (\ref{k}). Quantity $Q$ is positive for all known
analytic EOSs, namely, equation (\ref{lagr2}) yileds $Q=m^{-2}/\left(
m^2+\chi \right) $, equation (\ref{lagr3}) yileds $Q=4m_{*}^4/\left(
m_{*}^2-2\chi \right) ^4$, equation (\ref{lagr4}) yileds $Q=10\left(
m_{*}^2+\chi \right) ^2/m_{*}^8$, equation (\ref{lagr5}) yields $Q=3/m_{*}^4$%
. Hence, in the light of (\ref{evol}), the current increases behind the
shock wave 
\begin{equation}
\Delta \chi =\chi _{+}-\chi _{-}>0  \label{ev00}
\end{equation}
In other words, any stable discontinuity of the current is associated with a
shock wave propagating in that direction which corresponds to positive
increment of $\chi $.

Equations (\ref{w1})-(\ref{w2}), (\ref{evol}) determine the following
relationship between the velocities: $\left| w_{-}\right|
>c_{-}>c_{+}>\left| w_{+}\right| $ in the magnetic regime and $c_{+}>\left|
w_{+}\right| >\left| w_{-}\right| >c_{-}$ in the electric regime. Detailed
calculation according to formulas (\ref{w1})-(\ref{w2}) is shown in Fig.~\ref
{s} and \ref{ss}. All EOSs of bosonic strings (\ref{lagr2})-(\ref{lagr4})
reveal similar behavior.

\section{Transition between two regimes}

Consider a transition from the electric to magnetic regime. Let us imagine
it in the form of a joint shock composed of two consequent shocks (Fig.~\ref
{flo1}c). Let the first electric shock propagates along the string $\chi
_{-}<\chi _e<0$ and velocities $w_{-}$ and $w_e$ its determined according to
formula (\ref{w2}). Let the second magnetic shock propagates immediately
after the first shock, $\chi _{+}>\chi _m>0$ and velocities $w_{+}$ and $w_m$
are determined according to formula (\ref{w1}). Both shocks merge into a
composite shock in the limit $\chi _e\rightarrow $ $\chi _m\rightarrow 0$.
Let the flows before and behind the shock are $D_{-}<1$ and $D_{+}<D_{-}<1$,
respectively. Altogether we determine velocities $w_{-}=-D_{-}$ and 
\begin{equation}
w_{+}=\frac{D_{+}-D_{-}}{1-D_{+}D_{-}}  \label{d-1}
\end{equation}
in the co-moving reference frame (Fig.~\ref{flo1}b). Equations (\ref{w1})-(%
\ref{w2}) yield $w_e\rightarrow -1$ and $w_m\rightarrow -1$, while the flow
behind the first electric shock is 
\begin{equation}
D_e=\frac{w_{-}-w_e}{1-w_ew_{-}}\rightarrow 1  \label{d0}
\end{equation}
in the laboratory reference frame (Fig.~\ref{flo1}a). The flow behind the
second magnetic shock is 
\begin{equation}
\bar D_{+}=\frac{w_{+}-w_m}{1-w_mw_{+}}\rightarrow 1  \label{d2}
\end{equation}
in the reference frame comoving $D_e$ and altogether it is 
\begin{equation}
D_{+}=\frac{\bar D_{+}+D_e}{1+\bar D_{+}D_e}\rightarrow 1  \label{d00m}
\end{equation}
in the the laboratory reference frame. Therefore, the velocity behind the
second shock $D_{+}$ exceeds the shock wave velocity $D_{-}$ that implies
impossibility of a joint shock with a transition from the electric to
magnetic regime.

The limiting chiral regime $\chi =0$\ plays the role of a barrier that
cannot be ''tunneled'' by a single shock, and the latter cannot be a
switcher between magnetic and electric regimes of the current. The similar
repulsive character of chiral current $\chi =0$ is observed in the
equilibrium solution of the field-theoretic vorton model \cite{BS2008} where
no transition between the regimes is possible: the current vanishes $\chi
\rightarrow 0$ when the loop radius tends to infinity, however, the sign of $%
\chi $ is not changed. The shocks obey the same strange property: it is not
able to pass through the chiral point, maintaining continuous link between
the two regimes. It should be also noted that exact chiral limit $\chi =0$
is no more than a theoretical possibility: as soon as it is achieved, the
shock becomes asymptotically unstable at finite time interval because
formulas (\ref{w1})-(\ref{w2}) give $w=c=1$ rather than strict inequality (%
\ref{evol}) which warrants stability at arbitrary time interval.

\section{Conclusion}

\begin{figure}[tbp]
\caption{The ratio of shock to sound velocity $M_-=w^2_-/c^2_-$ (left) and $%
M_+=w^2_+/c^2_+$ (right) vs increment $\Delta \chi$ in the magnetic regime
of bosonic currents at $m=m_*=1$ at initial $\chi_-=0.1$ are plotted for
four models of EOS: dot - (\ref{lagr2}), dash - (\ref{lagr3}), dashdot - (\ref
{lagr4}), solid - (\ref{lagr5}). }
\label{s}{\includegraphics[scale=0.4]{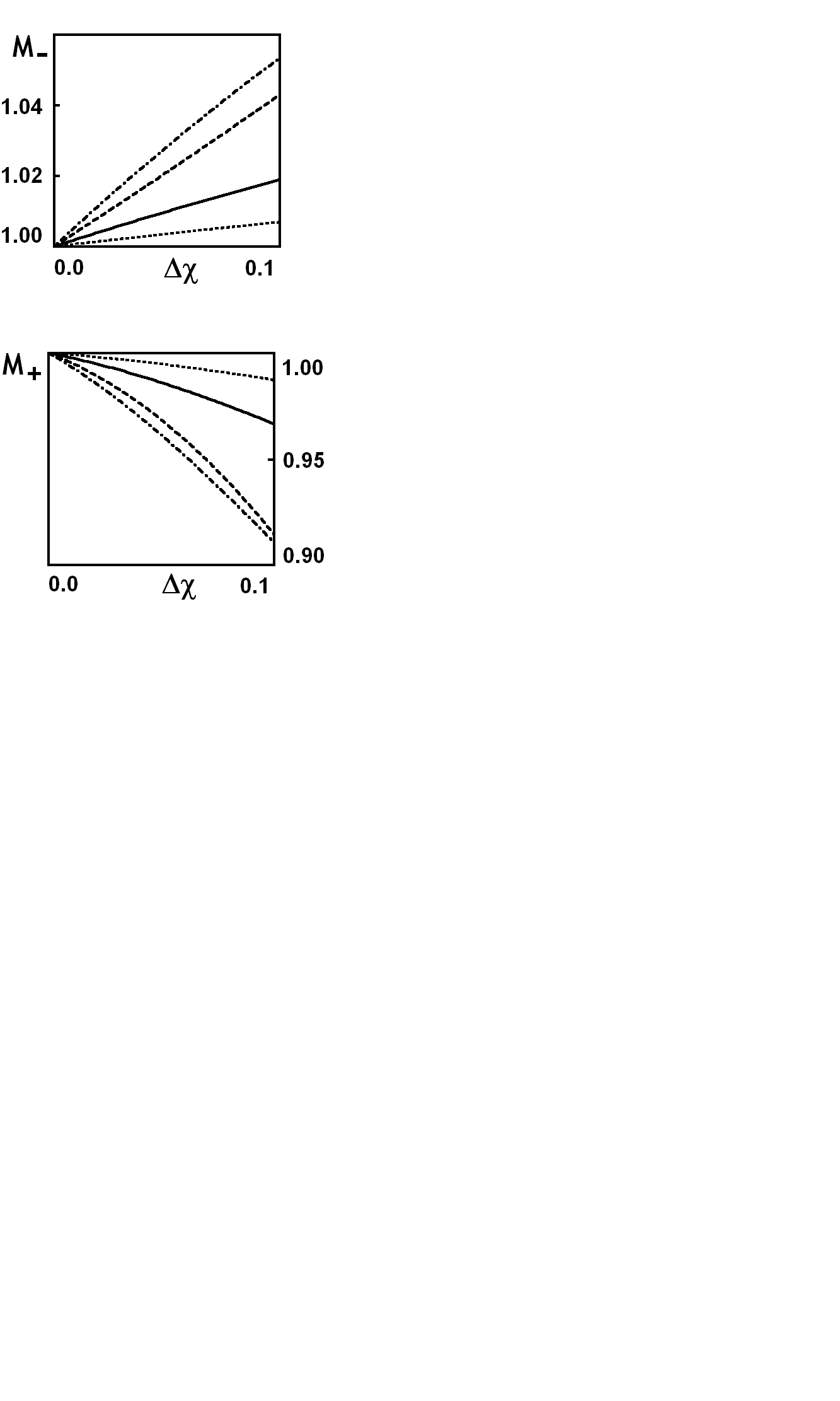}}
\end{figure}

\begin{figure}[tbp]
\caption{The same plots as in Fig.~(\ref{s}) but in the electric regime of
bosonic currents and at initial $\chi_-$=-0.05.}
\label{ss}{\includegraphics[scale=0.4]{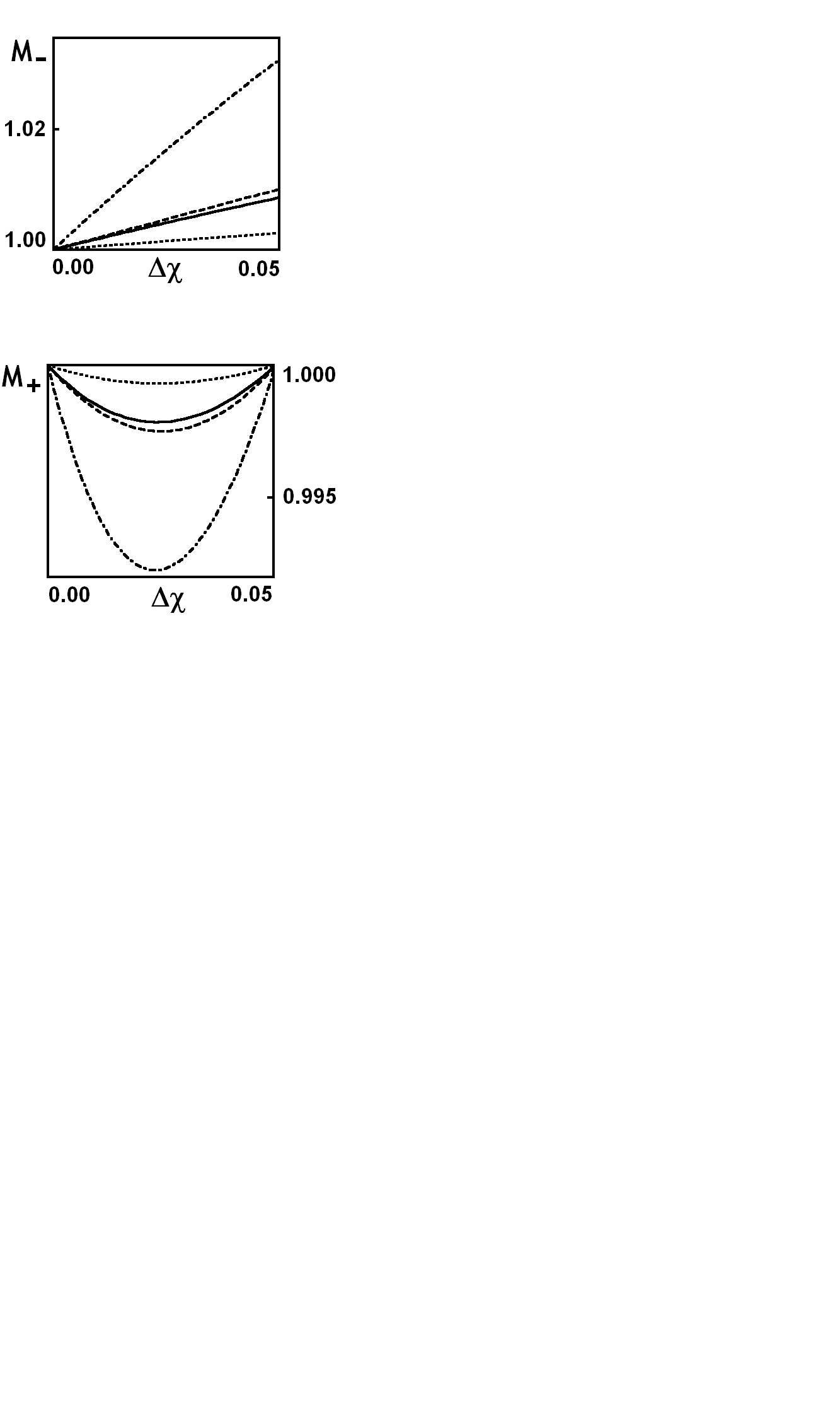}}
\end{figure}

The bosonic superconducting strings admit stable discontinuities (shock
waves) of the current. The shock velocity at space-like currents $\chi >0$\
(''magnetic'' regime) and time-like currents $\chi <0$\ (''electric''
regime) is determined by formulas (\ref{w1}) and (\ref{w2}), respectively.
Small-amplitude shocks are approximated by formulas (\ref{w11}) and (\ref
{w22}). Shock waves of arbitrary magnitude are calculated in Fig.~\ref{s}
and Fig.~\ref{ss}. No shock can provide a transition between the electric
and magnetic regimes, although electric shock is not forbidden by constraint
(\ref{ev00}). Numerical simulations \cite{MP00,CCMP02} does not reveal
existence of electric shocks at all, and the physical reason is hidden in
their vulnerability to transversal oscillations of the string worldsheet 
\cite{TV2011b} that is a problem of special investigation beyond the present
study.

The shock propagation implies increase of the current magnitude $\chi
_{+}>\chi _{-}$ (\ref{ev00}). It is similar to increase of the entropy in
ordinary shock waves because no stable discontinuity is admitted if the
current decreases. Condition (\ref{ev00}) is equivalent to $K_{+}>K_{-}$ for
all known models (\ref{lagr2})-(\ref{lagr4}). According to (\ref{x1}) it
implies $n_{+}>n_{-}$ in the magnetic regime. The latter inequality is
immediately recognized in numerical simulations where a shock structure is
associated with abrupt jump from lower $n_{-}$ to higher $n_{+}$ (see
especially Fig.~10 in Ref. \cite{MP00}).

In practice, shock waves can appear during non-linear evolution of small
longitudinal perturbations in the intrinsic equations of motion (\ref{cur1}%
). Intercommutation of two string loops with different currents $\chi _{-}>0$
and $\chi _{+}>0$ may also trigger a shock. Reconnection of two loops with $%
\chi _{+}<0$ and $\chi _{-}<0$ cannot trigger a shock \cite{MP00,CCMP02}.
Reconnection of two loops with opposite signs of the current (say, $\chi
_{-}<0$ and $\chi _{+}>0$) will result in appearance of the chiral state $%
\chi =0$ which must remain the same without regard of the radius \cite
{BS2008}. Although the loop radius must increase and tend to infinity as
soon as $\chi \rightarrow 0$, neither smooth nor shock transition from the
electric to magnetic regime is possible. We may expect an irreducible point
of inflection at $\chi =0$ and zero curvature (infinite radius), dividing
the electric and magnetic sides of the string, while the shock itself can
exist at the magnetic side. This problem deserves more attention in
application to scenarios of possible observable phenomena.

We have considered an ideal stationary problem with no external force. The
real cosmic string networks may include time-dependent parameters and
numerical analysis. The analytic: Whenever a shock wave is formed, the
current is subject to increase (\ref{ev00}), and this general law may give
help in the further theoretical and experimental research of superconducting
strings.

We wish to thank Alexei Starobinsky and Konstantin Stepanyantz for support.

\newpage

\end{document}